\newtheorem{definition}{Definition}
\newtheorem{theorem}{Theorem}
\newcommand{\bt}{\begin{theorem}}
\newcommand{\et}{\end{theorem}}
\newcommand{\bd}{\begin{definition}}
\newcommand{\ed}{\end{definition}}
\newcommand{\be}{\begin{equation}}
\newcommand{\ee}{\end{equation}}
\newcommand{\bear}{\begin{eqnarray}}
\newcommand{\eear}{\end{eqnarray}}
\newcommand{\baar}{\begin{array}}
\newcommand{\eaar}{\end{array}}
\newcommand{\nn}{\nonumber}
\newcommand{\pr}{\partial}
\newcommand{\G}{\Gamma}

\documentclass[11pt]{article}

\usepackage{latexsym}
\usepackage{amsfonts}
\usepackage[dvips]{graphicx}

\begin{document}
\renewcommand{\thesection}{\Roman{section}.}
\renewcommand{\thesubsection}{\Alph{subsection}.}
\thispagestyle{empty} 
\setlength{\textheight}{250mm}
\setlength{\topmargin}{-20mm}
\begin{center}
\large{\bf The Weyl-Lanczos Equations and the Lanczos Wave Equation in 4 
       Dimensions as Systems in Involution}
\end{center}
\vspace{0.5cm}\noindent
P Dolan$^{\: a)}$\\
Mathematics Department, Imperial College, 180 Queen's Gate,\\
London SW7 2BZ
\newline
\vspace{0.5cm}
\newline
A Gerber$^{\: b)}$\\
Centre for Techno-Mathematics and Scientific Computing Laboratory,\\
University of Westminster, Watford Road, Harrow HA1 3TP
\vspace{0.5cm}\indent
\begin{center} 
\large{Abstract} 
\end{center}

Using the work by Bampi and Caviglia, we write the Weyl-Lanczos 
equations as an exterior differential system. Using Janet-Riquier theory,
we compute the Cartan characters for all spacetimes with a diagonal metric 
and for the plane wave spacetime since all spacetimes have a plane wave limit.

We write the Lanczos wave equation as an exterior differential system 
and, with assistance from Janet-Riquier theory, we find  that it forms
a system in involution. This result can be derived from the scalar wave 
equation itself. We compute its Cartan characters and compare them 
with those of the Weyl-Lanczos equations.
\vspace{0.2cm}
\footnoterule
\vspace{0.2cm}\noindent
$^{a)}$ Electronic mail: pdolan@inctech.com\\
$^{b)}$ Electronic mail: a\_gerber01@hotmail.com
\newpage
\setlength{\textheight}{250mm}
\setlength{\topmargin}{-20mm}
\section{Introduction}
\subsection{The Weyl-Lanczos equations and the Lanczos Tensor Wave Equation
in 4 Dimensions}
Lanczos \cite{Lanc} generated the spacetime Weyl conformal tensor $C_{abcd}$ 
from a tensor potential $L_{abc}$ by covariant differentiation such that 
$C_{abcd}$ is given by
\bear
C_{abcd} & = &
L_{abc;d}-L_{abd;c}+L_{cda;b}-L_{cdb;a}
+g_{bc}L_{(ad)}+g_{ad}L_{(bc)}-g_{bd}L_{(ac)} \nn\\
& & -g_{ac}L_{(bd)}+\frac{2}{3}{L^{ms}}_{m;s}(g_{ac}g_{bd}-g_{ad}g_{bc}) 
\: ,\label{1Weyll}
\eear
where $L_{(ad)}={{L_{a}}^{s}}_{d;s}-{{L_{a}}^{s}}_{s;d}$ and $'';''$ denotes
covariant differentiation. We call (\ref{1Weyll}) the Weyl-Lanczos equations. 
The index symmetries of the Lanczos tensor $L_{abc}$ have to match the 
symmetries of (\ref{1Weyll}) and so it is usual to impose
\be
L_{abc} = L_{[ab]c} \: , \label{1bug}
\ee
and
\be
L_{[abc]} = 0   \label{1cyclic}
\ee
and the trace free (gauge) condition
\be
{L_{a}}^{s}_{\: s}=0 \: . \label{1trace}
\ee
The spacetime Weyl Lanczos equations (\ref{1Weyll}) can also be expressed 
in the more compact form
\be
C_{abcd} =L_{[ab][c;d]} + L_{[cd][a;b]}- ^{*}L^{*}_{[ab][c;d]} 
- ^{*}L^{*}_{[cd][a;b]}\label{1WeylDD}
\ee
as done in \cite{Rob1}, where $``;''$ denotes covariant differentiation. The 
algebraic equations (\ref{1bug}), (\ref{1cyclic}),
(\ref{1trace}) leave us with only 16 independent components for the $L_{abc}$.
If we then introduce the differential gauge conditions
\be
{L_{ab}}^{s}_{\: ;s} = 0 \label{1diffg} \: ,
\ee
we can simplify (\ref{1Weyll}) considerably to get
\be
C_{abcd} = L_{abc;d}-L_{abd;c}+L_{cda;b}-L_{cdb;a}-g_{bc}L^{s}_{ad;s}
-g_{ad}L^{s}_{bc;s}+g_{bd}L^{s}_{ac;s}+g_{ac}L^{s}_{bd;s} \: .
\label{1WeylS}
\ee
Theoretically, we could completely solve the 6 differential gauge conditions 
for 6 further components and have 10 components for 10 independent spacetime 
Weyl-Lanczos equations. But this approach does not exhibit the most
general solution. We note that equations (\ref{1diffg}), (\ref{1WeylS}) 
constitute a system of linear first-order partial differential equations
in 4 dimensions which can easily be rewritten as an exterior differential 
system EDS in involution. This theory only applies in 4 dimensions
as the Weyl Lanczos problem in 2 and 3 dimensions does not exist and for 5 
dimensions, we expect extra conditions to apply as we shall point out 
in a later paper. It was already shown in \cite{Bam1} that the above 
4-dimensional Weyl-Lanczos equations consist of an exterior differential 
system (EDS) in involution with respect to the spacetime variables. One can 
obtain the same results by applying the corresponding Janet-Riquier theory 
\cite{Ger1}. 

From the Weyl-Lanczos problem it is possible to generate a tensor wave
equation for the (spacetime) Lanczos potential from which the Penrose wave 
equation for the Weyl tensor $C_{abcd}$ can be derived \cite{Dol1}. Arising
from the Weyl-Lanczos equations is the linear tensor wave equation
\be
\Box
L_{abc}+2R_{c}^{\: s}L_{abs}-R_{a}^{\: s}L_{bcs}-R_{b}^{\: s}L_{cas}
-g_{ac}R^{ls}L_{lbs}+g_{bc}R^{ls}L_{las}-{\small{\frac{1}{2}}}
R L_{abc}=J_{abc} \: , \label{Waff1}
\ee
where
\be
J_{abc} = \frac{1}{2}R_{c[a;b]}-\frac{1}{6}g_{c[a}R_{;b]} \label{Schout}
\ee
and
\be
   \Box L_{abc}=g^{sm}L_{abc;s;m} \: , \label{Waff2}
\ee
from which Penrose's non-linear wave equation\footnotemark[1] for the 
spacetime Weyl tensor
\bear
\Box C_{abcd} -{C_{ab}}^{\: \: sm}C_{smcd}+4C_{asm[c}{C^{m}_{\: \:\: d]}}^{s}
_{\: \: b}+ \frac{R}{4} C_{abcd} & = & J_{[ab][c;d]} + J_{[cd][a;b]} \\
& & - ^{*}J^{*}_{[ab][c;d]} -^{*}J^{*}_{[cd][a;b]} \nn
\eear
was derived in \cite{Pen2}.

It was shown by Bampi and Caviglia \cite{Bam1} that the 5 dimensional
version of equation (\ref{1Weyll}) with the appropriate equations generalising
(\ref{1bug}), (\ref{1cyclic}), (\ref{1trace}), (\ref{1diffg}) constitutes a 
system of linear first order partial differential equations which is also an
EDS in involution. In this paper we are going to show that like the 
Weyl-Lanczos problem (\ref{1Weyll}) the spacetime Lanczos tensor wave equation
(\ref{Waff1}) can be rewritten as an EDS in involution, a result which is 
confirmed using Janet-Riquier theory.
\section{The Weyl-Lanczos Equations in 4 Dimensions as a System in Involution}
In \cite{Bam1} it was shown that the Weyl-Lanczos relations are a system in 
involution with respect to the spacetime variables.
The Weyl-Lanczos equations always constitute of a system
in involution as opposed to the Riemann-Lanczos equations, even for vacuum
spacetimes when $C_{abcd}=R_{abcd}$, because each problem is based on 
different equations.
\subsection{The Weyl-Lanczos Equations as an EDS}
The theory of exterior differential systems (EDS) can be found in many places
such as in \cite{Cart,5man,Yang,Moli}. A review together with some results on 
the Riemann-Lanczos problem in 4 dimensions is given in \cite{DoGe1}.

The EDS for the Weyl-Lanczos equations, which form a
Pfaffian system, was already given in \cite{Bam1}. The Weyl-Lanczos equations 
together with the differential gauge condition (\ref{1diffg}) form a system 
of 16 linear first-order PDEs. In a local coordinate frame, these 16 
equations look like
\bear
f_{abcd} & = & C_{abcd} -P_{abcd} +P_{abdc} -P_{cdab} +P_{cdba}
+ \G^{n}_{ad}(L_{nbc}+L_{ncb}) \nn\\
& & - \G^{n}_{ac}(L_{nbd}+L_{ndb})
+\G^{n}_{bc}(L_{nad}+L_{nda})- \G^{n}_{bd}(L_{nac}+L_{nca}) \nn\\
& & +g_{bc}g^{ns}P_{nads} +g_{ad}g^{ns}P_{nbcs}-g_{bd}g^{ns}P_{nacs}
-g_{ac}g^{ns}P_{nbds} \nn\\
& & -g_{bc}g^{ns}(\G^{m}_{ns}L_{mad}+\G^{m}_{as}L_{nmd}
+\G^{m}_{ds}L_{nam})-g_{ad}g^{ns}(\G^{m}_{ns}L_{mbc} \nn\\
& &+\G^{m}_{bs}L_{nmc}+\G^{m}_{cs}L_{nbm})+g_{bd}g^{ns}(\G^{m}_{ns}
L_{mac} +\G^{m}_{as}L_{nmc}+\G^{m}_{cs}L_{nam}) \nn\\
& & +g_{ac}g^{ns}(\G^{m}_{ns}L_{mbd}+\G^{m}_{bs}L_{nmd}+\G^{m}_{ds}
L_{nbm}) \: , \nn\\
f_{ab} & = & g^{ns}(P_{abns}-\G^{m}_{as}L_{mbn}-\G^{m}_{bs}L_{amn}
        -\G^{m}_{ns}L_{abm}) \label{3WeyLoc} \: ,
\eear
where $f_{abcd}=0$ denotes the Weyl-Lanczos equations and $f_{ab}=
{L_{ab}}^{s}_{\: ;s}=0$ the differential gauge conditions in local 
coordinates. 
When we construct the corresponding EDS, we introduce the local coordinates 
$(x^{e},L_{abc},P_{abcd})$ on the jet bundle $\mathcal{J}^{1}(\mathbb{R}^{4},
\mathbb{R}^{16})$ which form our formal manifold $\mathcal{M}$ of formal 
dimension $N=4+16+64=84$.

The exterior derivatives of all equations in (\ref{3WeyLoc}) constitute 
our first 16 one-forms. We also have to add 16 {\bf contact conditions} 
$K_{abc}$ in order to make sure that the $P_{abcd}$ can be considered
as if they were the partial derivatives of the $L_{abc}$. 
These additional 16 one-forms are locally given by $K_{abc}=d L_{abc}
-P_{abce}d x^{e}$. In this way we obtain the Pfaffian system $\mathcal{P}$
\bear
d f_{abcd} & = & [C_{abcd,e}+\alpha_{abcde}+\gamma_{abcde}]d x^{e} 
- d P_{abcd} + d P_{abdc} - d P_{cdab} + d P_{cdba} \nn\\
& & +g_{bc}g^{ns}d P_{nads}+g_{ad}g^{ns}d P_{nbcs}-g_{bd}g^{ns}d P_{nacs}
-g_{ac}g^{ns}d P_{nbds} \nn\\
& & + \G^{n}_{ad}(d L_{nbc}+d L_{ncb}) - \G^{n}_{ac}(d L_{nbd}+d L_{ndb})
+ \G^{n}_{bc}(d L_{nad}+d L_{nda}) \nn\\
& & - \G^{n}_{bd}(d L_{nac}+d L_{nca})-g_{bc}g^{ns}(\G^{m}_{ns}d L_{mad}
+\G^{m}_{as}d L_{nmd}+\G^{m}_{ds}d L_{nam}) \nn\\
& & -g_{ad}g^{ns}(\G^{m}_{ns}d L_{mbc}+\G^{m}_{bs}d L_{nmc}+\G^{m}_{cs}
d L_{nbm})+g_{bd}g^{ns}(\G^{m}_{ns}d L_{mac} \nn\\
& & +\G^{m}_{as}d L_{nmc} +\G^{m}_{cs}d L_{nam})+g_{ac}g^{ns}(\G^{m}_{ns}
d L_{mbd}+\G^{m}_{bs}d L_{nmd} \nn\\
& & +\G^{m}_{ds}d L_{nbm}), \nn\\
d f_{ab} & = & (P_{abns}g^{ns}_{\: \: ,e}
        -L_{mbn}\G^{m}_{as,e}-L_{amn}\G^{m}_{bs,e}
        -L_{abm}\G^{m}_{ns,e})d x^{e}\nn\\
  &   & +g^{ns}(d P_{abns}-\G^{m}_{as}d L_{mbn}-\G^{m}_{bs}d L_{amn}
        -\G^{m}_{ns}d L_{abm}), \nn\\      
K_{abc} & = & d L_{abc}- P_{abce}d x^{e} \: , \label{3WEDS} 
\eear
where $\alpha_{abcde}$ and $\gamma_{abcde}$ are given in Appendix A. Now, a
Vessiot vector field for (\ref{3WEDS}), where 16 of the totally 64 $V_{abcd}$ 
are determined through $d f_{abcd}(V)=0, d f_{ab}(V)=0$ from (\ref{3WEDS}),
will be of the form
\bear
V & = & V^{e}\frac{\pr}{\pr x^{e}}}+V^{e}P_{abce}{\frac{\pr}{\pr L_{abc}}
+ V_{abcd}\frac{\pr}{\pr P_{abcd}} \: .
\eear
Because $K_{abc}(V)=0$ has to hold as well, we get $V_{abc}=V^{e}P_{abce}$ as 
used above. We can determine the associated 
system $\mathcal{A}(\mathcal{P})$ of (\ref{3WEDS}) which is given locally by
$$\{ d f_{abcd}, d {L_{ab}}^{s}_{\: ;s}, \omega^{e}, d P_{abcd} \} \: ,$$
where $\omega^{e}:=d x^{e}$ and where we only include 48 of the 64 
$d P_{abcd}$. This is because 16 of them 
can be expressed by means of the exterior derivatives of the Weyl-Lanczos 
relations and the differential gauge equations. 
We see that $\mbox{dim}(\mathcal{A}(\mathcal{P}))=84=c,$ the class of 
$\mathcal{P}$, so that 
$\mbox{dim}(\bar{\mathcal{D}})$=0, where $\bar{\mathcal{D}}$ is defined as
\be
\bar{\mathcal{D}} = \{ Y \in \mathcal{D} / d \theta^{\alpha}(X,Y,)=0 \:
\forall X \in \mathcal{D} \: \forall 1 \leq \alpha \leq s \} \: , 
\ee
where $s$ denotes to the total number of independent 1-forms $\theta^{\alpha}$ 
in our Pfaffian system $\mathcal{P}$.
If we choose $Y \in Char(\mathcal{P})$, we have to be able to express 
$$ Y \rfloor d K_{abc} = Y^{e}d P_{abce}- Y_{abce}d x^{e}$$ by means of 
a non-trivial linear combinations of the forms $K_{abc}, d f_{ab}, d f_{abcd}$
with linear multipliers $\lambda^{a'b'c'},$\\
$\lambda^{a'b'c'd'}, \lambda^{nm}$
\be
\lambda^{a'b'c'}K_{a'b'c'}+\lambda^{a'b'c'd'} d f_{a'b'c'd'} 
+ \lambda^{mn}d {L_{mn}}^{s}_{\: ;s} \: .
\ee
However, no such linear combination can exist because some of the $d P_{abcd}$
will fail to occur in one or other of the $d f_{abcd}$ or the $d f_{ab}$ and 
in this way we know that {\it no Cauchy characteristics can occur}.

Next, we want to compute the reduced Cartan characters by hand for this 
Pfaffian system $\mathcal{P}$ using the tableau matrix. First, we must 
complete $(d f_{abcd},d f_{ab},\omega^{e})$ so that it becomes a complete 
coframe on $\mathcal{M}$, say $(d f_{abcd},d f_{ab},K_{abc},\omega^{e},
\pi^{\Lambda})$, where we use the forms in (\ref{3WEDS}) as cobasis elements. 
Accordingly, we must add the 48 new cobasis elements 
\[ \baar{ll}
\pi^{\Lambda}\leftrightarrow & d P_{abc1},\quad\Lambda =1, \cdots ,16 \: , 
\nn\\
\pi^{\Lambda}\leftrightarrow & d P_{abc2},\quad\Lambda =17, \cdots ,32\: , 
\nn\\  
\pi^{\Lambda}\leftrightarrow & d P_{abc3},\quad\Lambda =33, \cdots ,48 \: ,
\eaar \]
where the ordering of the $P_{abcd}$ based on the indices $abc$ is given in
Appendix A and where $\Lambda$ is a collective index subject to Einstein's
summation convention and corresponding to the set of indices $abc$ . 
Here, we write $\omega^{1}=d x^{1},\: \omega^{2}=d x^{2},\: 
\omega^{3}=d x^{3}, \: \omega^{4}=d x^{4}$ to form the independence 
condition given by $\Omega = \omega^{1}\wedge \omega^{2} \wedge \omega^{3} 
\wedge \omega^{4}$. Using this notation we can express the 16 exterior 
derivatives of the contact conditions, where the $\alpha$ are arranged in 
the same way as the $L_{abc}$ in Appendix A, as
\be
d \theta^{\alpha} = A^{\alpha}_{\Lambda 1}\pi^{\Lambda} \wedge \omega^{1}
+A^{\alpha}_{\Lambda 2}\pi^{\Lambda} \wedge \omega^{2}
+A^{\alpha}_{\Lambda 3}\pi^{\Lambda} \wedge \omega^{3}
- d P_{abc4}\wedge \omega^{4} \label{3Splot} \: .
\ee
Further calculations are based on the assumption that we can express each 
of the 16 $d P_{abc4}$ in (\ref{3Splot}) as a distinct linear combination of
the $d f_{abcd}, \: K_{abc}, \: d {L_{ab}^{s}}_{\: ;s}$ and the 
$\omega^{e}$. Later on, it will be easy to verify that this is true for all 
spacetimes with diagonal metric tensor. The tableau matrices 
$A^{\alpha}_{\Lambda2}$,$A^{\alpha}_{\Lambda3}$ and $A^{\alpha}_{\Lambda4}$ 
can now easily be determined from (\ref{3Splot}), where we obtain the only 
non-vanishing components to be $A^{\alpha}_{\Lambda2}=A^{\alpha}_{\Lambda3}
=A^{\alpha}_{\Lambda4}=-1$ if $\alpha$ and $\Lambda$ refer to the same group
of indices $abc$. We do not need to determine $A^{\alpha}_{\Lambda4}$ 
explicitly as it does not contribute to the characters. This is because the 
sum of the ranks of the tableau matrices of increasing order cannot exceed 48
and so the terms in $A^{\alpha}_{\Lambda 4}$ do not contribute to the rank. In
this way, the matrix  
\[ \left( \baar{c}
A^{\alpha}_{\Lambda \: 1} \nn\\
A^{\alpha}_{\Lambda \: 2} \nn\\
A^{\alpha}_{\Lambda \: 3} \nn\\
A^{\alpha}_{\Lambda \: 4} 
\eaar \right) \]
is a 64x48-matrix which looks like:
\[ \left( \baar{lll}
A & 0 & 0\nn\\
0 & A & 0\nn\\
0 & 0 & A\nn\\
X_{1} & X_{2} & X_{3}
\eaar \right) \: , \]
where $X_{1},X_{2},X_{3}$ stand for $A^{\alpha}_{\Lambda 4}$ and each A is 
given by the 16x16-matrix $A:=- \mathbb{I}_{16}$. From this we can immediately
deduce that the reduced characters are $s_{1}'=s_{2}'=s_{3}'=16$ and 
$s_{4}'=0$. However, this method does not supply $s_{0}'$ but here clearly 
$s_{0}'=s=32$ which is simply the number of independent 1-forms in 
(\ref{3WEDS}) consisting of the 10 exterior derivatives of the Weyl-Lanczos 
equations, the 6 exterior derivatives of the differential gauge conditions and
the 16 contact conditions $K_{abc}$.
The reduced characters $(s_{0}',s_{1}',s_{2}',s_{3}',s_{4}')$ then are given
by $(32,16,16,16,0)$. When (\ref{3WEDS}) is pulled back onto the submanifold,
where $f_{abcd}=0$ and $f_{ab}=0$, we get $(16,16,16,16,0)$.

Lastly, we make an Ansatz for a transformation which absorbs the 
remaining torsion terms $B^{\alpha}_{ij}$. Torsion terms can arise when we
write each $d \theta^{\alpha}$ of a Pfaffian system $\mathcal{P}$ as
\be
d \theta^{\alpha} = A^{\alpha}_{\lambda i}\pi^{\lambda}\wedge \omega^{i} 
+ \frac{1}{2} B^{\alpha}_{ij} \omega^{i} \wedge \omega^{j} + \frac{1}{2}
C^{\alpha}_{\lambda \kappa}\pi^{\lambda} \wedge \pi^{\kappa} \: .
\ee
The only non-vanishing torsion terms here are those of type
$B^{\alpha}_{i4}$. In order to make them vanish, we must look for a 
transformation $\Phi$ of the form
\[ \baar{ll}
\pi^{\Lambda} \rightarrow & \pi^{\Lambda} 
+ p^{\Lambda}_{\: i} \omega^{i}
\: , \nn\\
 & \nn\\
B^{\alpha}_{ij} \rightarrow & \tilde{B}^{\alpha}_{ij}=
B^{\alpha}_{ij} + A^{\alpha}_{\Lambda i}p^{\Lambda}_{\: j}
-A^{\alpha}_{\Lambda j}p^{\Lambda}_{\: i} \: , \label{3Trans}
\eaar \]
where we choose the $p^{\Lambda}_{\: i}$ in such a way that
$\tilde{B}^{\alpha}_{ij}=0$. Such $p^{\Lambda}_{\: i}$ must be solutions to 
\be
0 = B^{\alpha}_{i4}-A^{\alpha}_{\Lambda 4}p^{\Lambda}_{\: i} \: ,
\ee
where we can choose $p^{\Lambda}_{\: 4}=0$ because the $B^{\alpha}_{44}$ always
vanish due to skew symmetry. For any such solution of the 
$p^{\Lambda}_{i}$ the torsion is absorbed and the system is in involution. 
This rather cumbersome calculation can be carried out using a REDUCE code 
in \cite{Ger1} based on the package EDS \cite{Har1}. We obtained the result 
that the Cartan characters coincide with their reduced counterparts
and that the torsion can be absorbed for a number of spacetimes such as
Schwarzschild, Kasner, the Debever-Hubaut class, G\"{o}del, the pp-wave 
spacetimes and conformally flat spacetimes.
\subsection{The Weyl-Lanczos Equations as a System of PDEs}
The theory in this section and for the Lanczos wave equation below is based on
a modernised version of Janet-Riquier theory \cite{Jan1,Jan2,Riqu}. In this
modernised form Janet-Riquier theory can be found in \cite{Pom2,Sei1} and, 
a review together with some results on the Riemann-Lanczos problems in 2 and 3
dimensions is given in \cite{Ger1,DoGe2}.

The 10 independent Weyl-Lanczos relations and the 6 differential gauge 
conditions in local coordinates are given by (\ref{3WeyLoc}). 
Using the first computer code in Appendix C in \cite{Ger1}, we can derive the
symbol $\mathcal{M}_{1}$ for any spacetime with diagonal metric
\be
d s^{2}= a_{1}{(d x^{1})}^{2} - a_{2}{(d x^{2})}^{2}- a_{3}{(d x^{3})}^{2}
- a_{4}{(d x^{4})}^{2} \: ,\label{3Diagm}
\ee 
where $a_{1},a_{2},a_{3},a_{4}$ depend on all spacetime variables.
In order to obtain a ranking, we replace the 4 components 
$L_{121},L_{131},L_{141},L_{122}$ and their partial derivatives by solving 
(\ref{1trace}) for them and we choose an ordering $\succ$ for the $P_{abce}$ 
in such a way that $P_{abc4} \succ P_{abc3} \succ P_{abc2} \succ P_{abc1}$ 
and then the sets of $P_{abce}$ ordered according to 
$\mathcal{R}^{(W,4)}_{\succ}$ 
for each $e = 1,2,3,4$, where $\mathcal{R}^{(W,4)}_{\succ}$ is given 
in Appendix A. This produces an orderly ranking and induces such a ranking 
amongst the {\it symbol variables} $V_{abcd}$ \cite{Sei1,Pom2}, where we can 
solve each equation for a different variable $V_{abc4}$ corresponding to a 
$P_{abc4}$. The number in brackets indicates the corresponding Weyl-Lanczos
equation or differential gauge condition given in Appendix A. Then, 
$\mathcal{M}_{1}$ is given in orthonomic form by
\bear 
\fbox{1} \quad V_{3434} & = & \frac{a_{4}}{a_{1}}V_{1331}
+\frac{a_{3}}{a_{1}}V_{1441}-\frac{a_{4}}{a_{2}}V_{2332}
-\frac{a_{3}}{a_{2}}V_{2442}+V_{3443} \nn\\
\fbox{2} \quad V_{2434} & = & \frac{a_{4}}{a_{1}}V_{1321}-\frac{a_{4}}{a_{2}}
V_{2322}+V_{2443}+V_{3442} \nn\\
\fbox{3} \quad V_{2424} & = & \frac{a_{4}}{a_{3}}V_{2323}+V_{2442}-\frac{a_{2}
a_{4}}{a_{1}a_{3}}V_{1331}-\frac{a_{2}}{a_{3}}V_{3443} \nn\\
\fbox{4} \quad V_{2334} & = & V_{2343}+\frac{a_{3}}{a_{2}}V_{2422}+V_{3432} 
-\frac{a_{3}}{a_{1}}V_{1421} \nn\\
\fbox{5} \quad V_{2324} & = & \frac{a_{2}}{a_{1}}V_{1431}+V_{2342}
-V_{2423}-\frac{a_{2}}{a_{3}}V_{3433} \nn\\
\fbox{6} \quad V_{1424} & = & \frac{a_{4}}{a_{3}}V_{1323}-\frac{a_{4}}{a_{3}}
V_{1233}+V_{1442} -\frac{a_{4}}{a_{3}}V_{2331} \nn\\
\fbox{7} \quad V_{1434} & = & \frac{a_{4}}{a_{2}}V_{1232}-\frac{a_{4}}{a_{2}}
V_{1322}+V_{1443}+\frac{a_{4}}{a_{2}}V_{2321} \nn\\
\fbox{8} \quad V_{1334} & = & V_{1343} +\frac{a_{3}}{a_{2}}V_{1422}
-\frac{a_{3}}{a_{2}}V_{2421}-\frac{a_{3}}{a_{2}}V_{1242} \nn\\
\fbox{9} \quad V_{1234} & = & V_{1243}+V_{1342}-V_{1432}-V_{2341}+V_{2431} 
\nn\\
\fbox{10} \quad V_{1324} & = & V_{1243}+V_{1342}-V_{1423}+V_{2431} \nn\\
\fbox{11} \quad V_{1244} & = & -\frac{a_{4}}{a_{3}}V_{2331}-V_{2441}
+\frac{a_{4}}{a_{3}}V_{1332}+V_{1442}-\frac{a_{4}}{a_{3}}V_{1233} \nn\\
\fbox{12} \quad V_{1344} & = & -\frac{a_{4}}{a_{2}}V_{1322}-\frac{a_{4}}{a_{3}}
V_{1333}+\frac{a_{4}}{a_{2}}V_{2321}-V_{3441} \nn\\
\fbox{13} \quad V_{1444} & = & -\frac{a_{4}}{a_{2}}V_{1422}-\frac{a_{4}}{a_{3}}
V_{1433}+\frac{a_{4}}{a_{2}}V_{2421}+\frac{a_{4}}{a_{3}}V_{3431} \nn\\
\fbox{14} \quad V_{2344} & = & -\frac{a_{4}}{a_{1}}V_{1231}+\frac{a_{4}}{a_{1}}
V_{1321}-\frac{a_{4}}{a_{2}}V_{2322}-\frac{a_{4}}{a_{3}}V_{2333} \nn\\
\fbox{15} \quad V_{2444} & = & -\frac{a_{4}}{a_{1}}V_{1241}+\frac{a_{4}}{a_{1}}
V_{1421}-\frac{a_{4}}{a_{2}}V_{2422}-\frac{a_{4}}{a_{3}}V_{2433} \nn\\
\fbox{16} \quad V_{3444} & = & -\frac{a_{4}}{a_{1}}V_{1341}+\frac{a_{4}}{a_{1}}
V_{1431}+\frac{a_{4}}{a_{2}}V_{2342}-\frac{a_{4}}{a_{2}}V_{2432} \nn\\
& & -\frac{a_{4}}{a_{3}}V_{3433} \: . \label{3WeyIs}
\eear
All 16 equations in (\ref{3WeyIs}) are of class 4 because all variables 
$x^{1},x^{2},x^{3},x^{4}$ are multiplicative variables for each equation so
that we obtain $\beta^{(1)}_{1}=0,
\beta^{(2)}_{1}=0,\beta^{(3)}_{1}=0,\beta^{(4)}_{1}=16$, where the upper index
$(k)$ in $\beta^{(k)}_{q}$ indicates the number of multiplicative variables 
and the lower index the order of the system of partial differential 
equations. For the sum $\sum^{4}_{k=1}\beta^{(k)}_{1} = 16$, and because we 
only have 16 equations, this means that we are already using coordinates 
which are $\delta$-regular. One must use $\delta$-regular coordinates, 
which are coordinates which gradually maximise $\beta^{(n)}_{q}$, then 
$\beta^{(n-1)}_{q}+ \beta^{(n)}_{q}$ and so on, to ensure the results 
obtained are intrinsic. 

Then, the Cartan characters are 
$\alpha^{(1)}_{1}=16,\alpha^{(2)}_{1}= 16,\alpha^{(3)}_{1}=16,
\alpha^{(4)}_{1}=0$. They are computed according to the formula 
\be
\alpha^{(k)}_{q} = m {{n + q -k -1} \choose {q -1}} - \beta^{(k)}_{q}
\ee 
of which details can be found in \cite{Sei1,Pom2,Ger1,DoGe2}.
Now, we must verify that our symbol is involutive.  Firstly, we prolong
(\ref{3WeyIs}) by differentiating each equation in $\mathcal{R}_{1}$ 
with respect to $x^{1},x^{2},x^{3},x^{4}$. This leads to $\mathcal{M}_{2}$ 
consisting of 64 equations. We find that we can rewrite the prolonged
symbol $\mathcal{M}_{2}$ in such a way that each equation contains a distinct
component $V_{abcde}$ given in Appendix A so that $r(\mathcal{M}_{2})=64$.
But we also have $\sum^{4}_{k=1} k \cdot \beta^{(k)}_{1} =64$ which means
that $\mathcal{M}_{1}$ is involutive, where further details on this 
calculation are given in \cite{Ger1}. In order to show that the system 
is formally integrable, we have to 
verify that the canonical projection of $\mathcal{R}_{2}$ from second to
first order $\pi^{2}_{1} (\mathcal{R}_{2})$ coincides with $\mathcal{R}_{1}$ 
itself, which means that $\pi^{2}_{1}(\mathcal{R}_{2})=\mathcal{R}_{1}$. But 
our system is a linear system of PDEs and the only way an integrability 
condition can arise, is, if in any of the prolonged equations all the
second-order partial derivatives $S_{abcde}$ defined by $S_{abcde}=L_{abc,de}$
can be eliminated completely. This is impossible because 
$r(\mathcal{M}_{2})=64$ and so formal integrability follows. We conclude
that we have verified by using Janet-Riquier theory that the Weyl-Lanczos 
equations together with the differential gauge condition form a system in 
involution for spacetimes with diagonal metric tensors.

If we wish to look at the symbol for a general spacetime, the calculations
become more cumbersome. Therefore, we decide to look at the plane-wave 
limit which all spacetimes possess, and we shall analyse this system 
instead. A good account of the plane-wave limit of spacetimes is given in 
\cite{Pen3}. There, a part of a properly embedded null geodesic $\gamma$ is 
taken and a corresponding procedure applied which leads to $W_{\gamma}$ - 
a plane wave limit. All spacetimes can locally be expressed using a line 
element 
\bear
d s^{2} & & = 2d x^{1} d x^{2} + a {(d x^{3})}^{2} + 2 b_{3} d x^{2} d x^{3} 
        + 2b_{4}d x^{2} d x^{4} - c_{33}{(d x^{3})}^{2} \nn\\
& & - 2c_{34}d x^{3} d x^{4} -c_{44}{(d x^{4})}^{2} \: , \label{3Plap}
\eear
where $a,b_{3},b_{4},c_{33},c_{34},c_{44}$ are functions of all
4 coordinates. When a plane-wave limit is taken, the metric
(\ref{3Plap}) becomes \cite{Pen3}
\be
d s^{2} = 2d x^{1}d x^{2} - C_{33}{(d x^{3})}^{2}-2C_{34}d x^{3} d x^{4} 
- C_{44}{(d x^{4})}^{2} \: ,
\label{3Plimit}
\ee
where $C_{33}, C_{34}$ and $C_{44}$ are arbitrary functions of $x^{1}$ only.
We can determine the symbol for the metric (\ref{3Plimit}) but we see
that for whatever ranking we choose, the coordinates are not 
$\delta$-regular. This means that we would have to prolong the system
to second order and see whether we can obtain the desired result for
the prolonged system. But in order to avoid this, we perform a linear 
coordinate transformation \cite{Pom2} which we choose to be
\bear
d {\tilde{x}^{1}} & = & a_{11}d x^{1} + a_{12}d x^{2} \: ,\nn\\
d {\tilde{x}^{2}} & = & a_{21}d x^{1} + a_{22}d x^{2} \: , \nn\\
d {\tilde{x}^{3}} & = & d x^{3} \: ,\nn\\
d {\tilde{x}^{4}} & = & d x^{4} \: ,
\eear
where $a_{11},a_{12},a_{21},a_{22}$ are arbitrary constants. To simplify 
matters, we first look at the special case 
$a_{11}=a_{12}=a_{21}=\frac{1}{\sqrt{2}},a_{22}=-\frac{1}{\sqrt{2}}$, which
produces a new, only slightly different metric line element for 
(\ref{3Plimit}) which is
\be
d s^{2} = {(d \tilde{x}^{1})}^{2} - {(d \tilde{x}^{2})}^{2} -C_{33}
{(d \tilde{x}^{3})}^{2}-2C_{34}d \tilde{x}^{3} d \tilde{x}^{4} 
- C_{44}{(d \tilde{x}^{4})}^{2} \: .
\label{3Plimnew}
\ee
Even though this seems a minor transformation, the change from {\bf 
characteristic} to {\bf non-characteristic coordinates} is necessary in order 
to obtain intrinsic results for the Cartan characters because the new 
coordinates $(\tilde{x}^{e})$ are $\delta$-regular coordinates. Then, we 
evaluate (\ref{3WeyLoc}) for the new line element (\ref{3Plimit}) and order 
the $P_{abcd}$ such that $P_{abc1} \succ P_{abc2} \succ P_{abc3} \succ 
P_{abc4}$ instead whereas we leave the ordering in each set $P_{abce}$ 
unchanged as in $\mathcal{R}^{(W,4)}_{\succ}$ which leads to an orderly 
ranking. We can now solve each of the symbol equations for one distinct 
variable $V_{abc1}$ and so obtain an orthonomic system 
\bear
V_{1211} & = &
\frac{1}{\Delta}(V_{1222}\Delta+V_{1233}C_{44}-V_{1234}C_{34}
-V_{1243}C_{34}+V_{1244}C_{33})\nn\\
V_{1311} & = &
\frac{1}{\Delta}(V_{1322}\Delta+V_{1333}C_{44}-V_{1334}C_{34}
-V_{1343}C_{34}+V_{1344}C_{33})\nn\\ 
V_{1411} & = & \frac{1}{\Delta}(-V_{1223}C_{34}{C_{33}^{-1}}\Delta
-V_{1224}\Delta +V_{1333}C_{33}^{-1}C_{34}C_{44}-V_{1334}C_{44}\nn\\
 & & -V_{1343}C_{33}^{-1}{C_{34}}^{2}+V_{1344}C_{34}+V_{1422}\Delta 
+ V_{1433}(C_{44}-C_{33}^{-1}{C_{34}}^{2}))\nn\\
V_{1221} & = &
\frac{1}{\Delta}(V_{1212}\Delta-V_{1313}C_{44}+V_{1314}C_{34}
+V_{1413}C_{34}-V_{1414}C_{33}\nn\\
 & & +V_{2323}C_{44}-V_{2324}C_{34}-V_{2423}C_{34}
+V_{2424}C_{33})\nn\\
V_{1331} & = &
\frac{1}{\Delta}(V_{1212}(C_{33}{C_{34}}^{3}-{C_{33}}^{2}C_{44})
+V_{1313}\Delta-V_{2323}{C_{34}}^{2}\nn\\
 & & +V_{2324}C_{33}C_{34}-V_{2332}\Delta
+V_{2423}C_{33}C_{34}-V_{2424}{C_{33}}^{2})\nn\\
V_{2321} & = & \frac{1}{\Delta}(V_{1223}\Delta -V_{1232}\Delta 
+V_{1322}\Delta+V_{1333}C_{44}-V_{1334}C_{34}-V_{1433}C_{34} \nn\\
& & +V_{1434}C_{33})\nn\\
V_{2331} & = & \frac{1}{\Delta}(V_{1222}(C_{33}-{C_{33}}^{2}C_{44})
-V_{1233}\Delta -V_{1332}\Delta
+V_{1423}C_{33}C_{34} \nn\\
& & -V_{1424}{C_{33}}^{2}+ V_{1323}(C_{33}C_{44}-2{C_{34}}^{2})
+V_{1324}C_{33}C_{34})\nn\\
V_{2421} & = & 
\frac{1}{\Delta}(V_{1223}(C_{34}C_{44}-C_{33}^{-1}{C_{34}}^{3})
-V_{1242}\Delta +V_{1333}C_{33}^{-1}C_{34}C_{44}\nn\\
 & & -V_{1334}C_{44}+V_{1343}C_{33}^{-1}\Delta +V_{1422}\Delta
-V_{1433}C_{33}^{-1}{C_{34}}^{2}
+V_{1434}C_{34})\nn\\
V_{1231} & = & 
\frac{1}{\Delta}(V_{1213}\Delta +V_{1312}\Delta -V_{2322}\Delta
+V_{2343}C_{34}-V_{2344}C_{33}-V_{2433}C_{34} \nn\\
& & +V_{2434}C_{33})\nn\\
V_{1321} & = & 
\frac{1}{\Delta}(V_{1213}\Delta +V_{1312}\Delta
+V_{2333}C_{44}-V_{2334}C_{34}-V_{2433}C_{34}+V_{2434}C_{33})\nn\\
V_{1241} & = & 
\frac{1}{\Delta}(V_{1214}\Delta +V_{1412}\Delta +V_{2343}C_{44}
-V_{2344}C_{34} -V_{2422}\Delta \nn\\
& & +V_{2434}(C_{34}-C_{44})) \nn\\
V_{1421} & = &
\frac{1}{\Delta}(V_{1213}(C_{34}C_{44}-C_{33}^{-1}{C_{34}}^{3})
+V_{1412}\Delta +V_{2333}C_{33}^{-1}C_{34}C_{44} \nn\\
& & -V_{2334}C_{44}+V_{2343}C_{33}^{-1}\Delta-V_{2433}C_{33}^{-1}{C_{34}}^{2}
+V_{2434}C_{34})\nn\\
V_{1341} & = & 
\frac{1}{\Delta}(-V_{1212}C_{34}\Delta -V_{1314}\Delta
-V_{2323}C_{33}C_{44} +V_{2324}{C_{34}}^{2} +V_{2423}C_{33} \nn\\
& & C_{44}-V_{2424}C_{33}C_{34} -V_{2432}\Delta)\nn\\
V_{1431} & = & 
\frac{1}{\Delta}(-V_{1212}C_{34}\Delta +V_{1413}\Delta
-V_{2323}C_{34}C_{44} +V_{2324}C_{33}C_{44} -V_{2342}\Delta\nn\\
 & & +V_{2424}{C_{34}}^{2}-V_{2424}C_{33}C_{34})\nn\\
V_{2341} & = & 
\frac{1}{\Delta}(-V_{1222}C_{34}\Delta -V_{1234}\Delta
-V_{1323}C_{34}C_{44} +V_{1324}C_{33}C_{44}\nn\\ 
 & & +V_{1423}C_{33}C_{44}-V_{1424}C_{33}C_{34} -V_{1432}\Delta)\nn\\
V_{2431} & = & 
\frac{1}{\Delta}(-V_{1222}C_{34}\Delta -V_{1243}\Delta -V_{1323}C_{34}C_{44}
+V_{1324}C_{33}C_{44}-V_{1342}\Delta \nn\\
 & & +V_{1423}C_{33}C_{44} -V_{1424}C_{33}C_{34}) \: ,
\eear
where $\Delta = C_{33}C_{44}-{C_{34}}^{2}$. This system is composed of 
16 equations all of class 4 which leads to $\beta^{(1)}_{1}=\beta^{(2)}_{1}
=\beta^{(3)}_{1}=0,\beta^{(4)}_{1}=16$ and therefore to the Cartan 
characters $\alpha^{(1)}_{1}=\alpha^{(2)}_{1}=\alpha^{(3)}_{1}=16$ and
$\alpha^{(4)}_{1}=0$. This result is intrinsic because our coordinates are 
now $\delta$-regular.

Next, we prolong each of the 16 equations and obtain 64 equations
which can be modified so that each of the corresponding symbol equations 
contains a distinct variable $V_{abc11},V_{abc22},V_{abc33},V_{abc44}$ or 
$V_{abc34}$ and therefore $r(\mathcal{M}_{2})=64$. This agrees with 
$\sum_{k=1}^{4} k\cdot\beta^{(k)}_{1}=64$ so that $\mathcal{M}_{1}$ is 
involutive. No integrability conditions can occur for the same reasons as for
the system (\ref{3WeyIs}) above and the system is in involution with Cartan 
characters $(16,16,16,0)$. 

We obtain the same result for the plane wave limit of any 
spacetime as for the for the case of spacetimes with diagonal metric and we
conjecture that the same system on spacetimes with arbitrary analytic metric
will be in involution with Cartan characters $(16,16,16,0)$.
\section{The Lanczos Wave Equation in 4 Dimensions}
In this section, we look at the Lanczos wave equation at first as an
EDS and then as a system of PDEs. We again determine the Cartan characters and
show that it consists of a system in involution using both theories, a result 
which can be derived from the scalar wave equation directly.
\subsection{The Lanczos Wave Equation as an EDS}
We can also describe the Lanczos tensor wave equation in
terms of an EDS on a jet bundle $\mathcal{J}^{2}(\mathbb{R}^{4},
\mathbb{R}^{16})$ with formal dimension $N=244$ for which we choose 
the local coordinates $(x^{e},L_{abc},P_{abcd},S_{abcde})$ composed by
4 spacetime coordinates $x^{e}$, 16 $L_{abc}$, 64 $P_{abcd}$ and 256 
$S_{abcde}$. Here, again $S_{abcde}$ are the variables corresponding to the 
second-order partial derivatives of the $L_{abc}$ when projected onto our 
spacetime manifold. 

First, we write the Lanczos wave equation (\ref{Waff1}) in solved form and 
denote its components by $W_{abc}$. In a 
local coordinate frame $W_{abc}$ is then given by
\bear
W_{abc} & = &
  g^{sm}[S_{abcms}-\G^{n}_{am,s}L_{nbc}-\G^{n}_{bm,s}L_{anc}
-\G^{n}_{cm,s}L_{abn}-\G^{n}_{as}P_{nbcm} \nn\\
& & +\G^{n}_{as}\G^{k}_{nm}L_{kbc}+\G^{n}_{as}\G^{k}_{bm}L_{nkc}
+\G^{n}_{as}\G^{k}_{cm}L_{nbk}-\G^{n}_{bs}P_{ancm}
+\G^{n}_{bs}\G^{k}_{am} \nn\\
& & L_{knc} +\G^{n}_{bs}\G^{k}_{nm}L_{akc}+\G^{n}_{bs}\G^{k}_{cm}L_{ank}
-\G^{n}_{cs}P_{abnm}+\G^{n}_{cs}\G^{k}_{am}L_{kbn} \nn\\
& & +\G^{n}_{cs}\G^{k}_{bm}L_{akn}+\G^{n}_{cs}\G^{k}_{nm}L_{abk}
-\G^{n}_{ms}P_{abcn}+\G^{n}_{ms}\G^{k}_{an}L_{kbc} \nn\\
& & +\G^{n}_{ms}\G^{k}_{bn}L_{akc}+\G^{n}_{ms}\G^{k}_{cn}L_{abk}]
+2R_{c}^{\: s}L_{abs}-R_{a}^{\: s}L_{bcs}-R_{b}^{\: s}L_{cas} \nn\\
& & -g_{ac}R^{ls}L_{lbs}+g_{bc}R^{ls}L_{las}-\frac{1}{2}R L_{abc} -J_{abc}
\label{3WaveLoc} \: . 
\eear
In addition to the exterior derivatives of (\ref{3WaveLoc}), we need to add 
two sets of contact conditions, $K_{abc}$ and $K_{abcd}$, when we write 
(\ref{3WaveLoc}) as an EDS. Altogether, we obtain the Pfaffian system
\bear
d W_{abc} & = & d W_{abc} \nn\\
K_{abc} & = & d L_{abc}-P_{abce}d x^{e} \nn\\
K_{abcd} & = & d P_{abcd}-S_{abcde}d x^{e} \: , \label{3WAVEDS} 
\eear
where $d W_{abc}$ are the exterior derivatives of the components of
the wave equation in solved form, which are locally given by
\bear
d W_{abc} & = & [\Box L_{abc} + 2 R_{c}^{\: s}L_{abs}-R_{a}^{\: s}L_{bcs}
       -R_{b}^{\: s}L_{cas}-g_{ac}R^{ls}L_{lbs}+g_{bc}R^{ls}L_{las}\nn\\
& & -\frac{1}{2}RL_{abc}-J_{abc}]_{,e} d x^{e}+ g^{sm}[d S_{abcms}
-\G^{n}_{am,s}d L_{nbc}-\G^{n}_{bm,s}d L_{anc} \nn\\
& & -\G^{n}_{cm,s}d L_{abn}-\G^{n}_{as}d P_{nbcm}
+\G^{n}_{as}\G^{k}_{nm}d L_{kbc}+\G^{n}_{as}\G^{k}_{bm}d L_{nkc} \nn\\
& & +\G^{n}_{as}\G^{k}_{cm}d L_{nbk}-\G^{n}_{bs}d P_{ancm}
+\G^{n}_{bs}\G^{k}_{am}d L_{knc}+\G^{n}_{bs}\G^{k}_{nm}d L_{akc} \nn\\
& & +\G^{n}_{bs}\G^{k}_{cm}d L_{ank}-\G^{n}_{cs}d P_{abnm}
+\G^{n}_{cs}\G^{k}_{am}d L_{kbn}+\G^{n}_{cs}\G^{k}_{bm}d L_{akn} \nn\\
& & +\G^{n}_{cs}\G^{k}_{nm}d L_{abk}-\G^{n}_{ms}d P_{abcn}
+\G^{n}_{ms}\G^{k}_{an}d L_{kbc}+\G^{n}_{ms}\G^{k}_{bn}d L_{akc}\nn\\
& & +\G^{n}_{ms}\G^{k}_{cn}d L_{abk}]+2R_{c}^{\: s}d L_{abs}-R_{a}^{\: s}
d L_{bcs}-R_{b}^{\: s}d L_{cas}\nn\\
& & -g_{ac}R^{ls}d L_{lbs}+g_{bc}R^{ls}d L_{las}-\frac{1}{2}Rd L_{abc}\: . 
\eear
A Vessiot vector field for (\ref{3WAVEDS}) is of the form 
\be
V = V^{e}\frac{\pr}{\pr x^{e}}+V^{e}P_{abce}\frac{\pr}{\pr
L_{abc}}+V^{e}S_{abcde}\frac{\pr}{\pr P_{abcd}}+V_{abcde}
\frac{\pr}{\pr S_{abcde}} \: ,
\ee
where 16 of the 160 $V_{abcde}$ are determined by requiring 
that $d W_{abc}(V) =0$. When we apply the 2-forms $d K_{abc}, d K_{abcd}$
to the two Vessiot vector fields $V^{1}$ and $V^{2}$, 
we can see that $d x^{e} \wedge d P_{abce}(V^{1},V^{2})$ vanishes identically.
Therefore, we only have to examine the
second set $d K_{abcd}=-d x^{e} \wedge d S_{abcde}$ when we form our integral 
elements of dimensions greater than one. 

When we start to compute the Cartan characters of (\ref{3WAVEDS}), we obtain 
the values $s_{0}=s_{0}'=96$ again by counting the independent 1-forms in 
(\ref{3WAVEDS}) for the full and for the reduced system respectively.
These 1-forms consist of 16 exterior derivatives of the Lanczos wave equation, 
16 first-order contact conditions $K_{abc}$ and 64 second-order contact
conditions $K_{abcd}$ totalling 96 independent 1-forms.
If we pull the system back onto the submanifold defined by $W_{abc}=0$,
we obtain $s_{0}=s_{0}'=80$.

The associated system of (\ref{3WAVEDS}) can again be determined and it is
given by $$ \{ d W_{abc}, K_{abc}, K_{abcd}, \omega^{e},
d S_{abcde} \} \: , $$ 
where 16 of the totally 160 $d S_{abcde}$ can be expressed by
solving $d W_{abc}$ for them so that we only add the 144 remaining
$d S_{abcde}$ for which the $d W_{abc}$ are not solved for.
We see that $\mbox{dim} ( \mathcal{A}(\mathcal{P}))=244=c$ which 
means that $\mbox{dim}(\bar{\mathcal{D}})=0$ so that again, no Cauchy
characteristics can exist because for an $Y \in Char(\mathcal{P})$ 
$$ Y \rfloor d K_{abcd} = Y^{e}d S_{abcde}-Y_{abcde}d x^{e}$$ 
has to be such that its RHS can be expressed by means of a non-trivial
linear combination of the form
\be
\lambda^{a'b'c'}K_{a'b'c'}+\lambda^{a'b'c'd'}K_{a'b'c'd'}
+\lambda^{mns}d W_{mns} \: ,
\ee
which is impossible.

Again, we wish to determine the reduced Cartan characters using the tableau 
matrices $A^{\alpha}_{\lambda i}$ and examine whether the torsion terms 
$B^{\alpha}_{ij}$ of (\ref{3WAVEDS}) can really 
be absorbed. Therefore, we must complete $(d W_{abc},K_{abc},K_{abcd},
\omega^{i})$ to a coframe on our formally 244-dimensional jet-bundle. 
In (\ref{3WAVEDS}) we have 96 one-forms to which we add the 4 
one-forms $\omega^{1}=d x^{1}, \: \omega^{2}=d x^{2}, \: \omega^{3}=
d x^{3}, \: \omega^{4}=d x^{4}$ again composing the independence
condition $\Omega \neq 0$ so that we need to add a further 
144 forms $\pi^{\Lambda}$ in order to obtain a complete coframe of $N=244$ 
one-forms. Again, $\Lambda$ is a collective index again subject to Einstein's 
summation convention, this time replacing certain sets of indices $abcd$ in 
$d S_{abcde}$. In order to obtain intrinsic values, we choose to replace all 
components $d S_{abcde}$, where either $d$ or $e$ are $=4$, first, and so on 
based on $x^{4}$ being our first coordinate, then $x^{3}$, then $x^{2}$ and 
$x^{1}$ last. The correspondence of $\pi^{\Lambda} \: \leftrightarrow \: 
d S_{abcde}$ is given in detail in Appendix B. Note that we did not replace 
the 16 $d S_{abc11}$ as this would lead to an excess of 1-forms in our 
coframe but we solve each $d W_{abc}$ for a distinct $d S_{abc11}$.

Because $d K_{abc}=0 \: ( \mbox{mod} \: (\mathcal{P}))$, we only need to 
consider $d K_{abcd}$ when computing the tableau matrix and the reduced Cartan 
characters. We can now write the 64 $d K_{abcd}$ as
\be 
d \theta^{\alpha} \equiv A^{\alpha}_{\Lambda 2} \pi^{\Lambda}\wedge \omega^{2}
+A^{\alpha}_{\Lambda 3} \pi^{\Lambda}\wedge \omega^{3}
+A^{\alpha}_{\Lambda 4} \pi^{\Lambda}\wedge \omega^{4}
- d S_{abc11}\wedge \omega^{1} +B^{\alpha}_{ij}\omega^{i}\wedge \omega^{j} \: .
\ee
Here, we assumed that we can express all 16 $d
S_{abc11}$ by means of a distinct linear combination of the $d W_{abc}, \:
K_{abc}, \: K_{abcd}$ and the $\omega^{e}$.
From this we find that the only non-vanishing tableau matrix components 
$A^{\alpha}_{\Lambda i}$ for $i \neq 1$ are given by $A^{\alpha}_{\Lambda i}
=-1$ when the indices $\alpha$ and $\Lambda$ correspond to the same set of 
indices $abcd$ as given in Appendix B and $A^{\alpha}_{\Lambda i}=0$ 
otherwise. This leads to the tableau matrix
\[ \left( \baar{c}
A^{\alpha}_{\Lambda \: 4} \nn\\
A^{\alpha}_{\Lambda \: 3} \nn\\
A^{\alpha}_{\Lambda \: 2} \nn\\
A^{\alpha}_{\Lambda \: 1} 
\eaar \right) \]
which is a 160x144-matrix of the form\\
\[ M:= \left( \baar{lll}
0 & 0 & A\nn\\
0 & B & 0\nn\\
C & 0 & 0\nn\\
X_{1} & X_{2} & X_{3}
\eaar \right) \: , \]
where $X_{1},X_{2},X_{3}$ stand for $A^{\alpha}_{\Lambda 1}$ and A is a 
64x64-matrix of the form $-\mathbb{I}_{64}$, B a 
48x48-matrix $-\mathbb{I}_{48}$ and C a 32x32-matrix $-
\mathbb{I}_{32}$. We see that even without the 16 rows for the entries
$X$ the maximal rank $r(M)=144$ is obtained. This immediately leads to 
$s_{1}'=64$,$s_{2}'=48$, $s_{3}'=32$ and $s_{4}'=0$ meaning that the set of 
reduced Cartan characters $(s_{0}',s_{1}',s_{2}',s_{3}',s_{4}')$is 
$(96,64,48,32,0)$.

As a last step, we again give an Ansatz for a transformation $\Phi$ by means 
of which the torsion terms $B^{\alpha}_{ij}$ can be absorbed. We perform a 
transformation $\phi$ such 
that
\be
0 =
B^{\alpha}_{i1}-A^{\alpha}_{\Lambda 1}p^{\Lambda}_{\: i}, 
\quad i \neq 1 \; ,
\ee
and we set $p^{\Lambda}_{\: 1}:=1$ because all $B^{\alpha}_{ij}$ for 
both $i,j \neq 1$ vanish identically. This leads to $\tilde{B}^{\alpha}_{i1}
=0$ for any such solution for the $p^{\Lambda}_{\: i}$. 
Then, the torsion terms are absorbed and the system is in involution. Again, 
we do not carry out this longish calculation by hand but refer to results in
\cite{Ger1} confirming that the Lanczos wave equation consists of a system in 
involution with Cartan characters $(96,64,48,32,0)$. For a number of 
spacetimes such as Kasner, G\"{o}del,Schwarzschild, conformally flat 
spacetimes, this result was shown using the REDUCE code given in \cite{Ger1}.
\subsection{The Lanczos Wave Equation as a System of PDEs}
The Lanczos wave equation constitutes of 16 second-order equations
for the 16 independent Lanczos components, where we again imposed
(\ref{1cyclic}) and (\ref{1trace}). We can determine the symbol of
the wave equation and from this obtain the Cartan characters. 
We know that the wave equation has the form (\ref{Waff1}).
We denote the system formed by these 16 components of the wave equation by 
$\mathcal{R}_{2,wave}$, where the index ''2'' refers to the order of the 
system. Due to the definition of the symbol, only the highest-order 
derivatives matter, which are second-order derivatives here. Therefore, the 
only terms contributing to the symbol $\mathcal{M}_{2,wave}$ are parts of the 
$\Box L_{abc}$-terms, namely all the $g^{sm}\pr_{s}\pr_{m} L_{abc}$. The 16 
equations for the symbol for an arbitrary spacetime then look like
\bear
V_{abc44} & = & \frac{1}{g^{44}}(g^{11}V_{abc11} +g^{22}V_{abc22} 
+ g^{33}V_{abc33} +2g^{12}V_{abc12}+2g^{13}V_{abc13} \nn\\
& & +2g^{14}V_{abc14}+2g^{23}V_{abc23}+2g^{24}V_{abc24}
+2g^{34}V_{abc34}) \: . \label{3Wavsym}
\eear
We can easily see that by ordering the $S_{abcde}$ in such a way that 
$S_{abc44} \succ S_{abc34} \succ S_{abc33} \succ S_{abc24} \succ S_{abc23} 
\succ S_{abc22} \succ S_{abc14} \succ S_{abc13} \succ S_{abc12} \succ 
S_{abc11}$ based on $x^{4} \succ x^{3} \succ x^{2} \succ x^{1}$ and by 
choosing for each set of $S_{abcij}$ the ordering 
$\mathcal{R}^{(wave,4)}_{\succ}$ given in Appendix B, an orderly ranking is 
achieved.

Because all equations of form (\ref{3Wavsym}) are of class 4, it means 
that $\beta^{(4)}_{2}=16$ is the maximal
value for $\beta^{(4)}_{2}$. Because all $\beta^{(k)}_{2}=0$ for $k <
4$, the decreasing sums of the $\beta^{(k)}_{2}$ are automatically 16, 
which is the maximal value, so that the system is given in $\delta$-regular 
coordinates. Therefore, the following values for
$\beta^{(1)}_{2}=0,\beta^{(2)}_{2}=0,\beta^{(3)}_{2}=0,\beta^{(4)}_{2}=16$ 
lead to the intrinsic results for the Cartan characters which are
$\alpha^{(1)}_{2}=64,\alpha^{(2)}_{2}= 48,\alpha^{(3)}_{2}=32,
\alpha^{(4)}_{2}=0 \: .$ 

In order to see whether the system is in involution, we determine
$\mathcal{M}_{3,wave}$. Each component of the wave equation prolonged to 
third order contributes to the symbol $\mathcal{M}_{3,wave}$ with
\bear
0& = & g^{11}V_{abc111} +g^{22}V_{abc122} + g^{33}V_{abc133} 
+ g^{44}V_{abc144} +2g^{12}V_{abc112} \nn\\
& & +2g^{13}V_{abc113}+2g^{14}V_{abc114} +2g^{23}V_{abc123}
+2g^{24}V_{abc124}+2g^{34}V_{abc134} \nn\\
0 & = & g^{11}V_{abc112} +g^{22}V_{abc222} + g^{33}V_{abc233} 
+ g^{44} V_{abc244} +2g^{12}V_{abc122} \nn\\
& & +2g^{13}V_{abc123}+2g^{14}V_{abc124} +2g^{23}V_{abc223}
+2g^{24}V_{abc224}+2g^{34}V_{abc234} \nn\\ 
0 & = & g^{11}V_{abc113} +g^{22}V_{abc223} + g^{33}V_{abc333} 
+ g^{44} V_{abc344} +2g^{12}V_{abc123} \nn\\
& & +2g^{13}V_{abc133}+2g^{14}V_{abc134} +2g^{23}V_{abc233}
+2g^{24}V_{abc234}+2g^{34}V_{abc334} \nn\\
0 & = & g^{11}V_{abc114} +g^{22}V_{abc224} + g^{33}V_{abc334} 
+ g^{44} V_{abc444} +2g^{12}V_{abc124} +2g^{13} \nn\\
& & V_{abc134}+2g^{14}V_{abc144} +2g^{23}V_{abc234}
+2g^{24}V_{abc244}+2g^{34}V_{abc344} \: ,
\eear 
where 320 distinct symbol variables $V_{abcdef}$ can occur corresponding to 
the 320 distinct third order partial derivatives which can occur. This leads 
to a 64x320-symbol matrix
\[ \left( \baar{cccc}
S & 0 & \cdots & 0\nn\\
0 & S & \cdots & 0\nn\\
\vdots & \vdots & \ddots & \vdots \nn\\
0 & 0 & \cdots & S
\eaar \right) \: , \]
where each of the 16 matrices S have the form of the 4x20-matrix
\setlength{\arraycolsep}{0.5mm}
{\small \[ \left( \baar{cccccccccccccccccccc}
g^{11} & 2g^{12} & 2g^{13} & 2g^{14} & g^{22} & 2g^{23} & 2g^{24} &
g^{33} & 2g^{34} & g^{44} & 0 & 0 & 0 & 0 & 0 & 0 & 0 & 0 & 0 & 0\nn\\
0 & g^{11} & 0 & 0 & 2g^{12} & 2g^{13} & 2g^{14} & 0 & 0 & 0 & g^{22}
& 2g^{23} & 2g^{24} & g^{33} & 2g^{34} & g^{44} & 0 & 0 & 0 & 0\nn\\
0 & 0 & g^{11} & 0 & 0 & 2g^{12} & 0 & 2g^{13} & 2g^{14} & 0 & 0 &
g^{22} & 0 & 2g^{23} & 2g^{24} & 0 & g^{33} & 2g^{34} & g^{44} & 0\nn\\
0 & 0 & 0 & g^{11} & 0 & 0 & 2g^{12} & 0 & 2g^{13} & 2g^{14} & 0 & 0 &
g^{22} & 0 & 2g^{23} & 2g^{24} & 0 & g^{33} & 2g^{34} & g^{44}
\eaar \right) . \]} 
From this we can easily deduce that the rank of $S$ is always 4 independently
of the choice of the metric tensor so that 
$r(\mathcal{M}_{3,wave})=4\cdot 16=64$ holds. But this is equal to
$\sum^{4}_{k=1}k\cdot\beta^{(k)}_{2}=4\cdot 16=64$ so that 
$\mathcal{M}_{2,wave}$ is involutive.
In order to show that $\mathcal{R}_{2,wave}$ is formally integrable, 
we need to show that for the canonical projection $\pi^{3}_{2}
(\mathcal{R}_{3,wave})=\mathcal{R}_{2,wave}$. But again, our system is linear,
and the only way an integrability condition can occur for an involutive symbol
derived from a system of linear PDEs, is, if one of the equations in 
$\mathcal{R}_{3,wave}$ can be modified in such a way that no third-order 
partial derivatives occur. Because $r(\mathcal{M}_{3,wave})=64$ is maximal, 
this is not possible and the Lanczos tensor wave equation forms a system 
in involution with Cartan characters $(64,48,32,0)$. 

We can also derive this result directly from the {\it scalar wave equation}
\be
\Box \Psi = 0  \: , \label{scall}
\ee
where $\Psi$ is our scalar component depending on $x^{1},x^{2},x^{3},x^{4}$. 
For the scalar wave equation we use a formal manifold $\mathcal{M}$ with 
$N=19$ formal dimensions of which a local basis is given by the 4 $x^{e}$, 
1 dependent variable $\Psi$, 4 $P_{e}$ and 10 $S_{ef}$ on our jet bundle 
$\mathcal{J}^{2}(\mathbb{R}^{4},\mathbb{R})$. When projected onto the 
4-dimensional spacetime manifold, $P_{e}$ are the first-order and $S_{ef}$ 
are the second-order partial derivatives of $\Psi$. Trivially, the symbol of 
a single equation is always involutive, no matter what the ranking we chose,
is, so that we obtain $\beta^{(1)}_{2}=\beta^{(2)}_{2}=\beta^{(3)}_{2}=0$ and 
$\beta^{(4)}_{2}=1$ based on the fact that a single
equation is always of class $n$. We then obtain the characters
$$\alpha^{(1)}_{2}=4 \: , \alpha^{(2)}_{2}=3 \: , \alpha^{(3)}_{2}=2 \: ,
\alpha^{(4)}_{2} =0 \: .$$
We further find that $r(\mathcal{M}_{3})=4$ and therefore $\mbox{dim}
(\mathcal{M}_{3})=16$ whereas  $r(\mathcal{R}_{3})=5$ and $\mbox{dim}
(\mathcal{R}_{3})=30$. We also find that $\mbox{dim}(\mathcal{R}_{2})=14$ and
because the symbol is linear it $$\mbox{dim}(\mathcal{R}^{(1)}_{2})= 
\mbox{dim}(\mathcal{R}_{3})- \mbox{dim}(\mathcal{M}_{3})=30-16=14 \: .$$ 
From this we conclude that $\mbox{dim}(\mathcal{R}_{2})= \mbox{dim}
(\mathcal{R}^{(1)}_{2})=14$ and, because the system is linear, no 
integrability conditions can occur and the scalar wave equation (\ref{scall}) 
consists of a system in involution with characters $(4,3,2,0)$. Note that in 
the case of the Lanczos wave equation, we are dealing with 16 such equations 
because we have 16 independent components $L_{abc}$ and find the 
correspondence between the two sets of characters is given by 
$$16 \cdot (4,3,2,0) = (64,48,32,0)\: .$$

Lastly, we wish to compare the characters of the Weyl-Lanczos equations with 
those of the Lanczos wave equation. Therefore, we need to prolongate the 
Weyl-Lanczos equations once to second order. But for an involutive system the 
characters of the symbol $\mathcal{M}_{q+r}$ of the prolonged system together 
with the 
$\beta^{(k)}_{q+r}$ are directly determined by \cite{Sei1,Pom2}
\bear
\alpha^{k}_{q+r}& = & \sum_{i=k}^{n}{r+i-k-1 \choose r-1}
\alpha^{(i)}_{q} \: , \nn\\
\beta^{k}_{q+r}& = & \sum_{i=k}^{n}{r+i-k-1 \choose
r-1}\beta^{(i)}_{q} \: ,
\eear
where in our case $n=4,m=16,q=1,r=1$, which leads to 
the result (48,32,16,0) for $(\alpha^{(1)}_{2},\alpha^{(2)}_{2},
\alpha^{(3)}_{2},\alpha^{(4)}_{2})$ for the prolonged 
Weyl-Lanczos equations. This shows that, based on the $\alpha^{(k)}_{2}$ for 
both systems, the Weyl-Lanczos equations are more restrictive than the Lanczos
wave equation with characters (64,48,32,0). The general solution of the 
Weyl-Lanczos equations only depends on 16 arbitrary functions of 3 variables 
whereas the general solution for the Lanczos wave equations contains 32 
arbitrary functions of 3 variables.
\section*{Conclusion}
We obtained the Cartan characters $(s_{0},s_{1},s_{2},s_{3},s_{4})$ for the
Weyl-Lanczos equations given as a Pfaffian system in involution which are
(32,16,16,16,0) and (16,16,16,16,0) when pulled back onto the submanifold where
the Weyl-Lanczos equations themselves vanish identically. These results were
obtained assisted by REDUCE codes based on the EDS package. The Cartan 
characters could be obtained for a diagonalized spacetime and for the plane 
wave limit taken when using Janet-Riquier theory and they are (16,16,16,0).

For the Lanczos wave equation, we showed that it also consists of a Pfaffian
system in involution and that its Cartan characters are given by 
(96,64,48,32,0) or by (80,64,48,32,0) when pulled back. We related the Cartan 
characters and involutiveness of the Lanczos wave equation to the corresponding
results for the scalar wave equation. We also found that the Lanczos wave 
equation is less restrictive than the Weyl-Lanczos equations allowing for 32 
arbitrary functions of 3 variables as opposed to only 16.
\section*{Acknowledgements}
Both authors wish to thank D Hartley and W M Seiler for valuable discussions 
as well as Prof L S Xanthis. A Gerber would like to thank the Swiss National 
Science Foundation (SNSF) and the Dr Robert Thyll-D\"{u}rr Foundation.
\renewcommand{\theequation}{A.\arabic{equation}}
\setcounter{equation}{0}
\section*{Appendix A: The Weyl-Lanczos Equations}
Here, we give the expressions for the $\alpha_{abcde}$ and the $\gamma_{abcde}$
occurring in the EDS for the Weyl-Lanczos equations given by (\ref{3WEDS}).
The quantity $\alpha_{abcde}$ is determined through
\bear
\alpha_{abcde} & = & 
\G^{n}_{ad,e}(L_{nbc}+L_{ncb})+\G^{n}_{bc,e}(L_{nad}+L_{nda})
-\G^{n}_{ac,e}(L_{nbd}+L_{ndb}) \nn\\
& & -\G^{n}_{bd,e}(L_{nac}+L_{nca}) \: .
\eear 
The quantity $\gamma_{abcde}$ is given by
\bear 
\gamma_{abcde} & = & 
g_{bc}[P_{nads}g^{ns}_{\: \: ,e}-L_{mad}(\G^{m}_{ns}g^{ns})_{,e}
-L_{nmd}(\G^{m}_{as}g^{ns})_{\: ,e}-L_{nam}(\G^{m}_{ds}g^{ns})_{,e}]\nn\\
& & +g_{ad}[P_{nbcs}g^{ns}_{\: \: ,e}-L_{mbc}(\G^{m}_{ns}g^{ns})_{,e}
-L_{nmc}(\G^{m}_{bs}g^{ns})_{\: ,e}-L_{nbm}(\G^{m}_{cs}g^{ns})_{,e}]\nn\\
& & -g_{bd}[P_{nacs}g^{ns}_{\: \: ,e}-L_{mac}(\G^{m}_{ns}g^{ns})_{,e}
-L_{nmc}(\G^{m}_{as}g^{ns})_{\: ,e}-L_{nam}(\G^{m}_{cs}g^{ns})_{,e}]\nn\\
& & -g_{ac}[P_{nbds}g^{ns}_{\: \: ,e}-L_{mbd}(\G^{m}_{ns}g^{ns})_{,e}
-L_{nmd}(\G^{m}_{bs}g^{ns})_{\: ,e}-L_{nbm}(\G^{m}_{ds}g^{ns})_{,e}]\nn\\
& & +g_{bc,e}[P_{nads}g^{ns}-L_{mad}(\G^{m}_{ns}g^{ns})
-L_{nmd}(\G^{m}_{as}g^{ns})-L_{nam}(\G^{m}_{ds}g^{ns})]\nn\\
& & +g_{ad,e}[P_{nbcs}g^{ns}-L_{mbc}(\G^{m}_{ns}g^{ns})
-L_{nmc}(\G^{m}_{bs}g^{ns})-L_{nbm}(\G^{m}_{cs}g^{ns})]\nn\\
& & -g_{bd,e}[P_{nacs}g^{ns}-L_{mac}(\G^{m}_{ns}g^{ns})
-L_{nmc}(\G^{m}_{as}g^{ns})-L_{nam}(\G^{m}_{cs}g^{ns})]\nn\\
& & -g_{ac,e}[P_{nbds}g^{ns}-L_{mbd}(\G^{m}_{ns}g^{ns})
-L_{nmd}(\G^{m}_{bs}g^{ns}) \nn\\
& & -L_{nbm}(\G^{m}_{ds}g^{ns})] \: . \label{3Weygam}
\eear
When introducing the collective index $\Lambda$ for the additional coframe 
elements $\pi^{\Lambda}$, we ordered the $L_{abc}$ according to
$$L_{133} < L_{144} < L_{123} < L_{132} < L_{124} < L_{142} < L_{134} < 
L_{143}$$ $$< L_{232} < L_{233} < L_{242} < L_{244} < L_{234} < L_{243}
< L_{343} < L_{344}$$
from which we deduce the ordering for the $\pi^{\Lambda} \leftrightarrow
d P_{abce}$ in such a way that the $d P_{abce}$ are arranged like the 
$L_{abc}$ above. In this way the $\pi^{\Lambda}$ are labelled according to 
the correspondence 
\[ \baar{lll}
\pi^{1} \leftrightarrow d P_{1331} \quad & , \: \cdots , \quad & \pi^{16} 
\leftrightarrow d P_{3441} \: , \nn\\
\pi^{17} \leftrightarrow d P_{1332} \quad & , \: \cdots , \quad & \pi^{32} 
\leftrightarrow d P_{3442} \: , \nn\\
\pi^{33} \leftrightarrow d P_{1333} \quad & , \: \cdots , \quad & \pi^{48} 
\leftrightarrow d P_{abc3} \: .
\eaar \]
When expressing the Weyl-Lanczos equations as a system of PDEs, we used the 
standard orderly ranking (if not stated otherwise explicitely) which is based 
on the ordering $P_{abc4} \succ P_{abc3} \succ P_{abc2} 
\succ P_{abc1}$. Then, amongst each set of the $P_{abci}$ we order them
according to
$$ P_{344i} \succ P_{343i} \succ P_{243i} \succ P_{234i} \succ P_{244i} \succ
P_{242i} \succ P_{233i} \succ P_{232i} $$ $$ \succ P_{143i} \succ P_{134i} 
\succ P_{142i} \succ P_{124i} \succ P_{132i} \succ P_{123i} \succ P_{144i} 
\succ P_{133i} \: .$$
This leads to an orderly ranking for the $P_{abcd}$ for the Weyl-Lanczos 
equations denoted by $\mathcal{R}^{(W,4)}_{\succ}$.

Sometimes, we also use the following labelling for the Weyl-Lanczos equations
and the differential gauge conditions:
\[ \baar{llll}
\fbox{1} \leftrightarrow f_{1212} \: , &
\fbox{2} \leftrightarrow f_{1213} \: , &
\fbox{3} \leftrightarrow f_{1313} \: , &
\fbox{4} \leftrightarrow f_{1214} \: , \nn\\
 & & & \nn\\
\fbox{5} \leftrightarrow f_{1314} \: , & 
\fbox{6} \leftrightarrow f_{1323} \: , &
\fbox{7} \leftrightarrow f_{1223} \: , &
\fbox{8} \leftrightarrow f_{1224} \: , \nn\\
 & & & \nn\\
\fbox{9} \leftrightarrow f_{1234} \: , &
\fbox{10} \leftrightarrow f_{1324} \: , &
\fbox{11} \leftrightarrow {L_{12}}^{s}_{\: ;s} \: , & 
\fbox{12} \leftrightarrow {L_{13}}^{s}_{\: ;s} \: , \nn\\
 & & & \nn\\
\fbox{13} \leftrightarrow {L_{14}}^{s}_{\: ;s} \: , &
\fbox{14} \leftrightarrow {L_{23}}^{s}_{\: ;s} \: , & 
\fbox{15} \leftrightarrow {L_{24}}^{s}_{\: ;s}\: , &
\fbox{16} \leftrightarrow {L_{34}}^{s}_{\: ;s} \: .
\eaar \]
Lastly, we give independent symbol components $V_{abcde}$ for each of the 64 
equations for the prolonged symbol $\mathcal{M}_{2}$ for the Weyl-Lanczos 
equations which are:
\[ \baar{|c|c|c|c|c|} \hline
& \mbox{i = 1} \qquad & \mbox{i = 2} \qquad & \mbox{i = 3} \qquad & 
\mbox{i = 4} \qquad \nn\\ 
& & & & \nn\\ \hline             
\pr_{i} f_{1212} & V_{14411} & V_{23322} & V_{34433} & V_{34344}\nn\\
\pr_{i} f_{1213} & V_{13211} & V_{34422} & V_{24433} & V_{24344}\nn\\
\pr_{i} f_{1313} & V_{13311} & V_{24422} & V_{23233} & V_{24244}\nn\\
\pr_{i} f_{1214} & V_{14211} & V_{34322} & V_{23433} & V_{23344}\nn\\
\pr_{i} f_{1314} & V_{14311} & V_{23422} & V_{24233} & V_{23244}\nn\\
\pr_{i} f_{1323} & V_{23311} & V_{14422} & V_{13233} & V_{14244}\nn\\
\pr_{i} f_{1223} & V_{23211} & V_{12322} & V_{14433} & V_{14344}\nn\\
\pr_{i} f_{1224} & V_{24211} & V_{12422} & V_{13433} & V_{13344}\nn\\
\pr_{i} f_{1234} & V_{23411} & V_{14322} & V_{12433} & V_{12344}\nn\\
\pr_{i} f_{1324} & V_{24311} & V_{13422} & V_{14233} & V_{13244}\nn\\
\pr_{i} {L_{12}}^{s}_{\: ;s} & V_{24411} & V_{13322} & V_{12333} 
& V_{12444}\nn\\
\pr_{i} {L_{13}}^{s}_{\: ;s} & V_{34411} & V_{13222} & V_{13333} 
& V_{13444}\nn\\
\pr_{i} {L_{14}}^{s}_{\: ;s} & V_{34311} & V_{14222} & V_{14333} 
& V_{14444}\nn\\
\pr_{i} {L_{23}}^{s}_{\: ;s} & V_{12311} & V_{23222} & V_{23333} 
& V_{23444}\nn\\
\pr_{i} {L_{24}}^{s}_{\: ;s} & V_{12411} & V_{24222} & V_{24333} 
& V_{24444}\nn\\
\pr_{i} {L_{34}}^{s}_{\: ;s} & V_{13411} & V_{24322} & V_{34333} 
& V_{34444} \nn\\ \hline 
\eaar \: \: . \]
Based on this we can show that $r(\mathcal{M}_{2})=64$, where the detailed 
calculation is given in \cite{Ger1}.
\renewcommand{\theequation}{B.\arabic{equation}}
\setcounter{equation}{0}
\section*{Appendix B: The Lanczos Wave Equation}
We also need to specify the correspondence between the $\pi^{\Lambda}$ and the
$d S_{abcde}$ for the Lanczos wave equation. For both $i,j$ fixed we assume 
that the $d S_{abcij}$ are ordered according to the ordering of the $L_{abc}$ 
in Appendix A. For the first 64 $\pi^{\Lambda}$, where we have used the fact 
that $x^{4} \succ x^{3} \succ x^{2} \succ x^{1}$ for our independent variables
in order to obtain intrinsic results, we obtain the labelling (for the pairs 
of indices $ij = 44, 34, 24, 14$)
\[ \baar{lll}
\pi^{1} \leftrightarrow d S_{13344} \: \quad & , \: \cdots \: , \quad & 
\pi^{16} \leftrightarrow d S_{34444} \: , \nn\\
\pi^{7} \leftrightarrow d S_{13334} \: \quad & , \: \cdots \: , \quad & 
\pi^{32} \leftrightarrow d S_{34434} \: , \nn\\
\pi^{33} \leftrightarrow d S_{13324} \: \quad & , \: \cdots \: , \quad & 
\pi^{48} \leftrightarrow d S_{34424} \: , \nn\\
\pi^{49} \leftrightarrow d S_{13314} \: \quad & , \: \cdots \: , \quad & 
\pi^{64} \leftrightarrow d S_{34414} \: .
\eaar \]
Then, the next 48 $\pi^{\Lambda}$ are given by (for $ij = 33, 23, 13$)
\[ \baar{lll}
\pi^{65} \leftrightarrow d S_{13333} \: \quad & , \: \cdots \: , \quad & 
\pi^{80} \leftrightarrow d S_{34433} \: , \nn\\
\pi^{81} \leftrightarrow d S_{13323} \: \quad & , \: \cdots \: , \quad & 
\pi^{96} \leftrightarrow d S_{34423} \: , \nn\\
\pi^{97} \leftrightarrow d S_{13313} \: \quad & , \: \cdots \: , \quad & 
\pi^{112} \leftrightarrow d S_{34413} \: .
\eaar \]
The last 32 $\pi^{\Lambda}$ are denoted as follows (for $ij = 22, 12$)
\[ \baar{lll}
\pi^{113} \leftrightarrow d S_{13322} \: \quad & , \: \cdots \: , \quad & 
\pi^{128} \leftrightarrow d S_{34422} \: , \nn\\
\pi^{129} \leftrightarrow d S_{13312} \: \quad & , \: \cdots \: , \quad & 
\pi^{144} \leftrightarrow d S_{34412} \: .
\eaar \]
We assumed that all the $d S_{abc11}$ can be solved for by the 16 exterior 
derivatives of the wave equation $d W_{abc}$ and expressed the $d S_{abc11}$
in this way.

An orderly ranking $\mathcal{R}^{wave,4}_{\succ}$ for the 160 second-order
partial derivatives $S_{abcde}$ for the Lanczos wave equation is based on the
ordering $$S_{abc44} \succ S_{abc34} \succ S_{abc33} \succ S_{abc24} 
\succ S_{abc23} \succ S_{abc22}$$ $$\succ S_{abc14} \succ S_{abc13} \succ 
S_{abc12} \succ S_{abc11}$$ and then amongst each set $S_{abcij}$ for 
$i,j$ fixed in the same way as for the $P_{abci}$ in Appendix A, namely:
$$S_{344ij} \succ S_{343ij} \succ S_{243ij} \succ S_{234ij} \succ S_{244ij}
\succ S_{242ij} \succ S_{233ij} \succ S_{232ij}$$ $$\succ S_{143ij} \succ
S_{134ij} \succ S_{142ij} \succ S_{124ij} \succ S_{132ij} \succ S_{123ij} 
\succ S_{144ij} \succ S_{133ij} \: .$$ 
\vspace{0.3cm}
\footnoterule
\vspace{0.2cm}\noindent
$^{1}$ The Penrose wave equation for $C_{abcd}$ can be found for all dimensions
$n \leq 4$. It was not derived from the Lanczos wave equation, which was given
for $n=5$ by Edgar and H\"{o}glund \cite{Idiot2}, and we postpone the 
discussion of the 5-dimensional Weyl-Lanczos problem and the related tensor 
wave equation to another paper.
\bibliography{refs}
\end{document}